\documentstyle[12pt,a4,psfig]{article}
\def\eg{{\it e.g.\ }}
\def\etal{{\rm et al.\thinspace}}
\def\HI{H{\sc i}}
\def\ie{{\it i.e.\ }}
\def\Lsun{\hbox{$\rm\thinspace L_{\odot}$}}
\def\Msun{\hbox{$\rm\thinspace M_{\odot}$}}
\def\reference{\parskip 0pt\par\noindent\hangindent 0.5 truecm}
\begin{document}

\title
{Cosmology using the Parkes Multibeam Southern-Sky \HI\ Survey}
\author
{P. A. Thomas\\
Astronomy Centre, CPES, University of Sussex\\
Falmer, Brighton, BN1$\,$9QH, UK\\
p.a.thomas@sussex.ac.uk}

\maketitle

\begin{abstract}
I discuss the implications of the Parkes \HI\ Multibeam Southern Sky
Survey for cosmology.  It will determine the local mass function of
\HI\ clouds, detecting several hundred per decade of mass.  Each of
these will come with a redshift and, for the more massive clouds, an
estimate of the velocity width.  This will provide an ideal database
for peculiar motion studies and for measurements of biasing of
galaxies relative to the underlying matter distribution.
\end{abstract}

Surveys --- Galaxies: luminosity function, mass function ---
large-scale structure of the Universe
\bigskip

In this paper, I wish to discuss the implications for cosmology of the
Parkes Multibeam Southern-Sky \HI\ Survey (hereafter referred to as
`the Multibeam Survey').  Most importantly, it will provide the first
unbiased survey for \HI\ clouds, either associated with galaxies or
truly isolated: I will show below that many thousand new objects
should be discovered whose mass-function will place great constraints
on models of galaxy formation.  The second important feature of the
Multibeam Survey is that it will map the velocity structure of the
local Universe in great detail.  When combined with accurate
Tully-Fisher distances, this will enable strong constraints to be
placed on cosmological parameters, principally the density parameter.

\section{The mass function of \HI\ clouds}

I will start by estimating the baryonic mass-function of collapsed
halos, assuming that baryons and dark matter are distributed equally
over the sky (\ie there is no biasing).  I will take parameters
appropriate for the standard Cold Dark Matter (CDM) cosmology (others
would give similar results): density parameter, $\Omega_0=1$; Hubble
parameter, $h=H_0/100\,$km$\,$s$^{-1}$Mpc$^{-1}=0.5$; baryon fraction,
$\Omega_b=0.064$; normalisation, $\sigma_8=0.6$.

An analytic estimate for the number density of halos as a function of
mass was first provided by Press \& Schechter (1974).  To estimate the
proportion of the Universe which is contained in structures of mass
$m$ at redshift $z$, the density-field is first smoothed with a
top-hat filter of radius $R$, where $m=(4\pi/3)\overline{\rho}R^3$ and
$\overline{\rho}$ is the mean density of the Universe.  $F(m,z)$ is
then defined to be the fractional volume where the smoothed density
exceeds some critical density $\delta_c$.  Assuming a gaussian 
distribution, then
\begin{equation}
F(m,z)={1\over 2}{\rm erfc}\left(\delta_c\over\sqrt{2}\sigma(m,z)\right),
\end{equation}
where $\sigma$ is the root-mean-square fluctuation within the top-hat
filter and ${\rm erfc}$ is the complementary error function.  The key
step was to realize that fluctuations on different mass-scales are not
independent.  In fact, to a first approximation Press \& Schechter
assumed that high-mass halos were entirely made up of lower-mass ones
with no underdense matter mixed in.  Then $F$ must be regarded as a
cumulative mass fraction and it can be differentiated to obtain the
fraction of the universe contained in structures of a given mass,
\begin{equation}
f(m,z)=-{\partial F\over\partial m}=-{1\over\sqrt{2\pi}}
{\delta_c\over\sigma^2}{\partial\sigma\over\partial m}
\exp\left\{-\delta_c^2/2\sigma^2\right\}.
\end{equation}
To convert this to a number density of halos per logarithmic
mass-interval we simply multiply by $\overline{\rho}$:
\mbox{d$n/$d$\ln m=\overline{\rho}f$}.
The main drawback of this approach is that, because of the above
assumption of crowding together of low-mass halos into larger ones, it
seems to undercount the number of objects.  However modern techniques
(\eg Bond \etal 1992) give the same analytic form, simply scaled by a
factor of two in normalization.  The modified formula gives good
agreement with numerical simulations (\eg Lacey \& Cole 1994) for
$\delta_c=1.69$ (as is appropriate for a spherical, top-hat collapse
of density peaks.

\begin{figure}
$$\vbox{
\psfig{figure=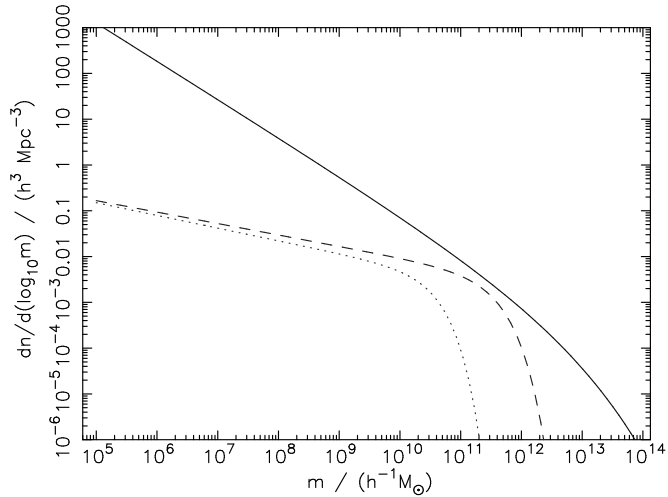,height=6.5truecm,width=8.7truecm,angle=270}
}$$
\caption{The mass function of halos in the standard CDM cosmology
(solid line) compared with the observed luminosity function of
galaxies (dashed line) assuming $m/L=15\,h\,\Msun/\Lsun$.  Also shown is
the \HI\ content of normal galaxies (dotted line).}
\label{fig:massfn}
\end{figure}

Figure~\ref{fig:massfn} shows the predicted mass function for the
baryonic content of halos and contrasts it with the observed stellar
mass function.  I have assumed a Schechter luminosity function,
\begin{equation}
{\rm d}N={{\cal L}\over L_*\Gamma(2+\alpha)}
\left(L\over L_*\right)^\alpha\exp\left\{-L/L_*\right\}
{\rm d}\left(L\over L_*\right),
\end{equation}
where ${\cal L}=1.7\times10^8h^3\Lsun$Mpc$^{-3}$,
$L_*=10^{10}h^{-2}\Lsun$, \mbox{$\alpha=-1.25$}, and the stellar
mass-luminosity ratio is $m_{\rm star}/L=15h$.

It is apparent that the model predicts approximately the correct
number density of normal ($L\sim L_*$) galaxies.  However, it gives far
too many halos at both higher and lower masses.  The reason for the
former discrepancy was first explained by Rees \& Ostriker (1977) and
by Silk (1977).  They compared the ratio of the cooling time of the
gas in proto-galactic halos to the dynamical time of the halo.  In
small objects the cooling time is always shorter than the dynamical
time, thus gas can cool to form stars and hence a visible galaxy.  In
larger systems, however, the cooling time exceeds the dynamical time.
Then mergers may shock-heat the gas before it has time to cool, thus
preventing significant star-formation.  For this reason clusters of
galaxies have little on-going star-formation except perhaps in a
cooling flow deep within the cluster core.  The dividing mass between
these two regimes is highly sensitive to the gas fraction and cooling
function but covers the range corresponding to the exponential cut-off
in the luminosity function.

The predicted excess of low-mass halos is harder to explain.  It would
seem that not all the cooled gas in proto-galaxies has formed visible
stars.  Until I researched this paper, it seemed possible to me that
the missing gas resided inside the halos of low-mass galaxies in the
form of \HI.  The \HI\ content of normal spiral galaxies is
insufficient (see, for example, the mass function shown by the dotted
line in Figure~\ref{fig:massfn} which is taken from the model in
Briggs 1990) but I thought it possible that there may be a significant
population of gas-rich dwarfs.  However, the observations of the
number counts of \HI\ clouds, described below, seem to rule this out.
This is consistent with the optical luminosity function which,
although it may miss many low-surface-brightness, predominantly
low-mass galaxies is unlikely to have a faint-end slope as steep as
the required value of $\alpha\approx -1.8$.  A more realistic
explanation for the missing gas is that much of the proto-galactic
interstellar medium was heated by an early generation of supernovae
(and/or strong galactic winds) and expelled from the halo.  If only a
small fraction of \HI\ remains, however, or if it has fallen back into
the galaxy, then it should be visible with the Multibeam Survey.

\begin{figure}
$$\vbox{
\psfig{figure=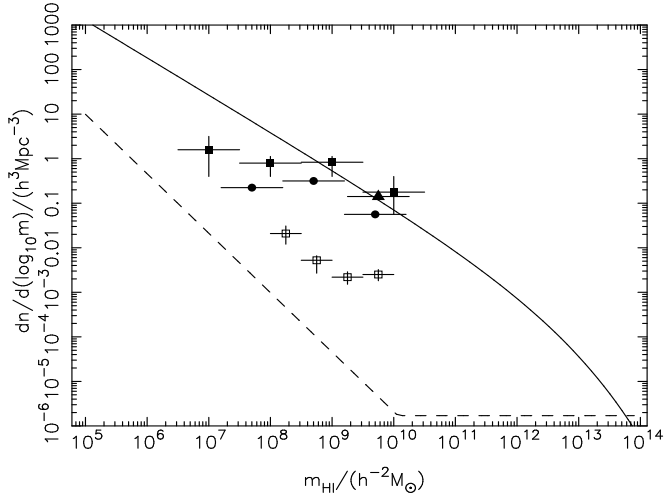,height=6.5truecm,width=8.7truecm,angle=270}
}$$
\caption{The detected density of \HI\ clouds (\ie galaxies) of
different mass, together with the expected sensitivity from the
Multibeam Survey.  For details see the text.}
\label{fig:obshi}
\end{figure}

Figure~\ref{fig:obshi} shows the observed number density of \HI\
clouds in various surveys.  The solid data points are taken from
Briggs (1990).  They all come from pointed observations towards
different clusters, but including different proportions of foreground
and background objects: solid squares, Leo group (Schneider \etal
1989); circles, Virgo (Hoffman \etal 1989); triangle, Hercules
(Salpeter \& Dickey 1985).  Unfortunately these surveys are all highly
biased.  The only blind survey for \HI\ clouds which I know of is that
of Kerr \& Henning (1987) which has recently been analysed by Henning
(1985).  That gives a lower space-density as shown by the open squares
in the figure.  A realistic estimate of the true space-density is
probably given by the lower locus of the points in the figure, some
one to two orders-of-magnitude below the predicted curve if all the
missing matter were in the form of \HI.  As the observations must be
shifted by about two orders-of-magnitude to the right to give good
agreement with the predictions, this suggests that about one percent
of the original baryonic mass of the halo persists in the form of \HI.

The dashed line in Figure~\ref{fig:obshi} shows the space density of
\HI\ clouds at which the Multibeam Survey will detect 10 objects
of a given mass.  The sensitivity is almost three orders-of-magnitude
higher than any previous survey thanks mainly due to the large volume
of space which is covered.  In making this prediction I have have
assumed that for a 5 sigma detection
\begin{equation}
m_{\rm\HI}\geq2.6\times10^7\Msun\,\left(D\over10\,{\rm Mpc}\right)^2
\left(\Delta v\over100\,{\rm km}\,{\rm s}^{-1}\right)^{0.5},
\end{equation}
where the data have been binned to match the velocity width, $\Delta
v$, of the cloud.  Supposing one per cent of the baryons to be in
the form of \HI, then the Tully-Fisher relation is
\begin{equation}
\Delta v\approx 440\,
\left(m_{\rm\HI}\over1.5\times10^9h^{-1}\Msun\right)^{0.22}
{\rm km}\,{\rm s}^{-1}.
\end{equation}
which gives a minimum density for the detection of $N$ clouds of
\begin{equation}
\phi\geq1.1\times10^{-3}\left(N\over10\right)
\left(m_{\rm\HI}\over10^8h^{-2}\Msun\right)^{-1.335}h^{0.165}
h^3\,{\rm Mpc}^{-3}.
\end{equation}
The flat part of the sensitivity curve arises for \HI\ clouds which
are detectable to the survey limit of 13$\,$500 km$\,$s$^{-1}$.

On the basis of Figure~\ref{fig:obshi}, I predict that the Multibeam
Survey will detect several hundred \HI\ clouds per decade of mass.  I
would expect most of these to be associated with dwarf galaxies, but
there may be a few truly intergalactic gas clouds.  The mass and
spatial distribution of these objects will be a major constraint on
models of galaxy formation.

\section{Numerical simulations of \HI\ clouds}

The formation of bound objects by the growth and collapse of small
density fluctuations in the early Universe is a highly complicated
process.  Although the Press-Schechter formalism, described above,
gives a good estimate of the number density of objects, it tells us
little about their spatial distribution or their internal structure.
For this we have to rely on N-body simulations.  Recently there has
been a great improvement in the power of such simulations resulting
primarily from two causes: firstly the introduction of massively
parallel computers consisting of a large number of processors
(typically a few hundred) each with their own memory and linked
together by high-speed data channels, and secondly the development of
sophisticated numerical algorithms able to take advantage of the new
machines.  Consequently pure N-body simulations (\ie gravity
only, or dark matter only) of a few tens of million particles and
N-body, hydrodynamical simulations (\ie a mixture of gas and dark
matter) of a few million particles are now practicable.

\begin{figure}
$$\vbox{
\psfig{figure=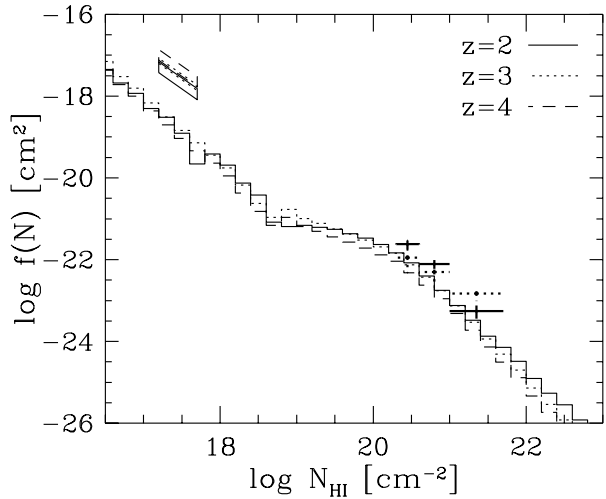,width=12truecm}
}$$
\vspace*{-8.5truecm}
\caption{A histogram of the frequency distribution of Ly$\alpha$
absorbing clouds per line-of-sight ($f(N)=$d$^2N/$d$N_{\rm\HI}$d$z$)
in the numerical simulation of Katz \etal (1995).  Error crosses and
diagonal boxes show the observational constraints from damped
Ly$\alpha$ and Lyman limit systems, respectively.}
\label{fig:nhi}
\end{figure}

Studies of absorption lines in quasar spectra show that the
high-redshift Universe is full of Ly$\alpha$ clouds, \ie clouds of
neutral hydrogen.  Katz \etal (1995) simulate the production of these
clouds in the standard CDM cosmology using parameters similar to those
described above.  They use 64$^3$ each of dark matter and gas
particles with a gas particle mass of $7.75\times10^7h^{-1}\Msun$.
The comoving volume of the box is $11.11\,h^{-1}$Mpc and the box is
evolved to a redshift of 2.  A uniform photo-ionizing background is
assumed to be present since a redshift of 6. Their paper contains a
beautiful picture of the surface density of neutral hydrogen at the
final time.  It shows a dense network of interconnected filamentary
structures studded with bright knots representing large concentrations
of cold gas.  They calculate the optical depth to \HI\ absorption as a
function of frequency for a variety of lines-of-sight through the box.
The resultant spectra are then analysed in a similar manner to quasar
spectra to produce a histogram of absorption-line equivalent-widths.
They are able to to reproduce the general form of the observed
distribution over a wide range of equivalent widths, from
$10^{14}$--$10^{22}$cm$^{-2}$.  The low-equivalent-width systems arise
from the filaments themselves and from velocity caustics of the gas
which is falling onto them.  However, the high-equivalent-width
systems, $N_{\rm\HI}\geq10^{17}$cm$^{-2}$, shown in
Figure~\ref{fig:nhi}, occur when the line-of-sight passes through a
lump of collapsed gas, \ie a galaxy (or more accurately a proto-galaxy
as there is no star formation in the code).  It can be seen from the
figure that the model predicts too few absorption-line systems.  This
may point to a deficiency in the standard CDM model, but is of little
concern to us here.

High-resolution simulations of this kind are very time-consuming and
it is impractical to carry them forward beyond a redshift of 2 to the
present day.  Nor would it be sensible to do so because of all the
uncertainties in the physics of the intergalactic medium.  However,
the Multibeam Survey will anyway only be sensitive to column densities
in excess of about $10^{18}$cm$^{-2}$ and we have just seen that these
are associated with galaxies whose distribution can be determined in
simulations of much lower resolution.  This is one of the goals of the
Virgo Consortium, a collaboration of mainly UK astronomers to carry
out cosmological N-body hydrodynamical simulations of the formation of
structure.  We have been awarded time on the Cray~T3D supercomputer at
Edinburgh that consists of 512 Dec~alpha chips (of which 256 are
usable at one time) each with 64$\,$Mb of memory.  We use the Hydra
code developed by Couchman, Thomas \& Pearce (1995) and available from
http://coho.astro.uwo.ca/pub/hydra/hydra.html.  Currently we are
carrying out dark matter simulations with 17 million particles and
dark matter plus gas simulations with 4 million particles.  Only the
latter are of relevance for this paper.

\begin{figure*}
$$\vbox{
\psfig{figure=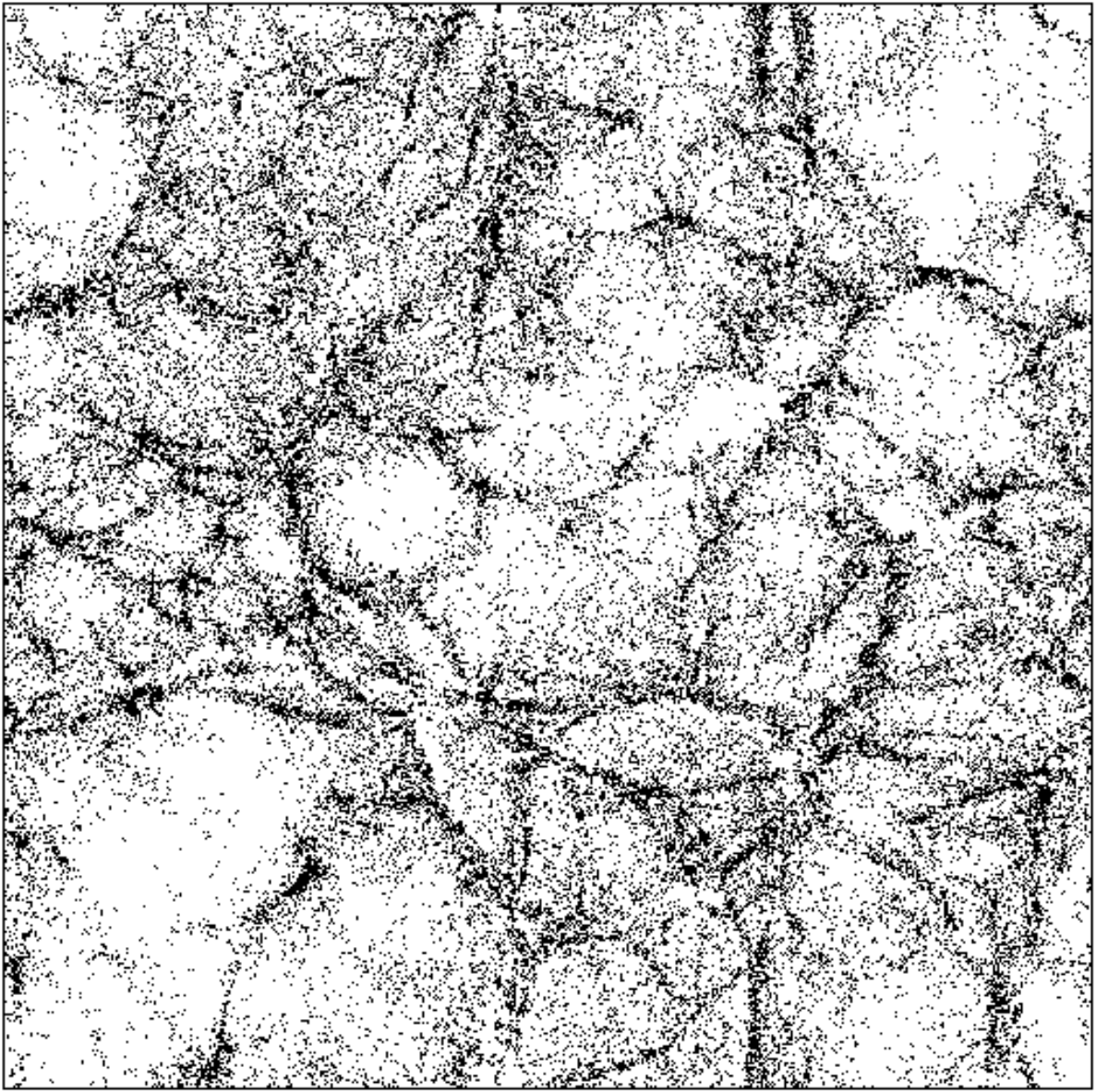,width=16truecm,angle=270}
\caption{The projected distribution of cold gas at $z=0$ in a slice through a
simulation of a critical-density, CDM universe (see text for
details).}
}$$
\label{fig:hicol}
\end{figure*}

Figure~\ref{fig:hicol} shows the distribution of cold gas ($T<10^5$K)
in one of our simulations at $z=0$.  It is a slice $7\,h^{-1}$Mpc
thick through a box $70\,h^{-1}$Mpc in width.  The cosmological
parameters are again similar to those given above, but the particle
mass is now $1.4\times10^9h^{-1}\Msun$.  The box-size is well-matched
to the size of the Multibeam Survey, although the mass-resolution is
poorer than one would like and allows us to model the distribution of
only the more massive galaxies.  However, I would expect an
improvement of a factor of eight in mass-resolution within a couple of
years.  If one looks carefully then one can see numerous lumps of cold
gas spaced along and at the intersection of filaments: these we
associate with galaxies.  It should be noted, however, that we have
deliberately kept the physics in these simulations to a minimum.  In
particular, we have not attempted to form stars with all the
associated feedback of energy into the interstellar medium.  For this
reason the \HI\ mass in the simulations is not representative of that
in real galaxies.

\begin{figure*}
$$\vbox{
\psfig{figure=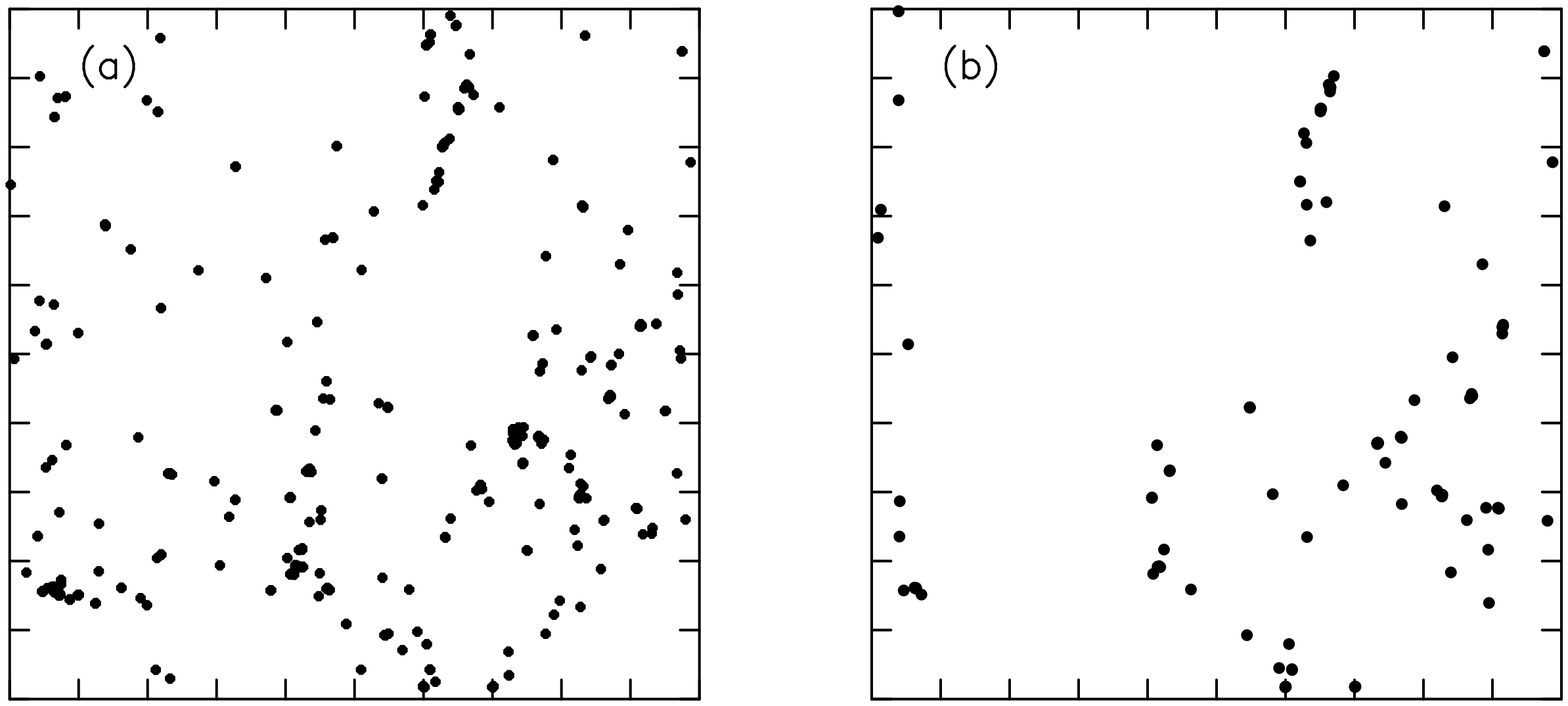,width=15truecm,angle=270}
}$$
\caption{Projections of the galaxy distribution in a test simulation:
(a) $m<0.5M_*$, (b) $m>0.5M_*$, where $M_*=1.5\times10^{11}h^{-1}\Msun$.}
\label{fig:proj}
\end{figure*}

One of the main purposes of including gas in the simulations is to
test the bias in the relative distributions of galaxies and dark
matter.  The former are more highly-correlated in space and are also
moving more slowly than the latter.  Moreover, the degree of
biasing is dependent upon the mass of the galaxies, being larger for
more massive systems.  As an illustration of this,
Figure~\ref{fig:proj} shows the relative distributions of moderate
and high-mass galaxies in a test simulation.  It will be very
interesting to see from the Multibeam Survey whether there is a large
population of dwarf galaxies filling the voids in the bright galaxy
distribution.  

The degree of biasing is a major bug-bear of cosmology because it
confuses the link between observations and theory: we see the galaxies,
but the models predict only the overall distribution of matter.   For
example, one of the best ways to estimate the density parameter,
$\Omega_0$, is to measure the peculiar velocities of galaxies relative
to the uniform Hubble expansion.  The expected motions are proportional
to $\Omega_0^{0.6}$ and also to the overdensity of matter, that is $1/b$
times the overdensity of light where $b$ is the bias parameter.
The Multibeam Survey will be an ideal database for peculiar motion
studies.  As well as detecting all large spiral galaxies out to more
than $100\,h^{-1}$Mpc, it will also measure their redshift and their
\HI\ velocity width.  When combined with infra-red photometry, this
latter quantity will enable accurate Tully-Fisher distances and hence
peculiar velocities.  

\begin{figure*}
\vspace*{-3truecm}
$$\vbox{
\psfig{figure=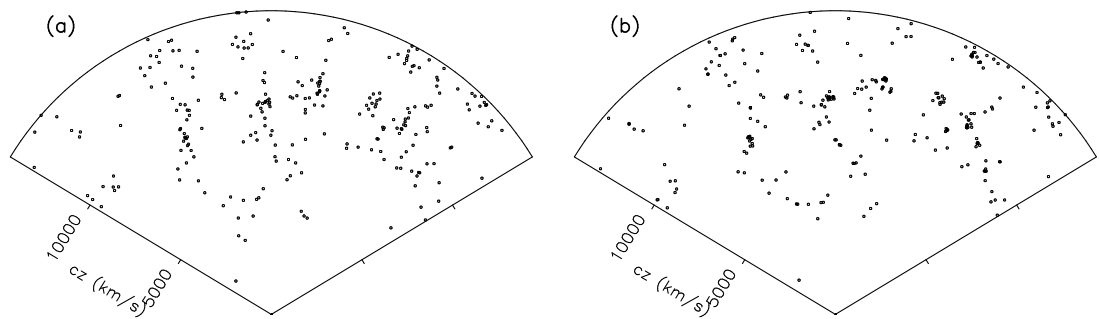,width=15truecm,angle=270}
}$$ 
\vspace*{-3truecm}
\caption{`Wedge diagrams' for the galaxy distribution in a simulation of
spatial extent similar to the Multibeam Survey: (a) redshift space, (b)
real space.}
\label{fig:slice}
\end{figure*}

Figure~\ref{fig:slice} shows `wedge diagrams' for the real and velocity
space distribution of galaxies drawn from an N-body simulation with
$\Omega=0.3$ and extent similar to the Multibeam Survey.  The high
density of galaxies which is expected in the Multibeam Survey provides
an advantage over other sparser surveys: if one can find a well-defined
void similar to those seen in the figure, then one can obtain a
constraint on $\Omega_0$ which is independent of the bias parameter. 
This is because the underdensity in a void can never be greater than
unity (whereas the overdensity in a cluster can in principle be
anything).

\section{Conclusions}
\begin{enumerate}
\item The Parkes Multibeam \HI\ Southern Sky Survey will determine the
mass function of \HI\ clouds in the local Universe.  This will provide
a strong constraint on models of galaxy formation.
\item The spatial distribution of \HI\ clouds of differing mass will
test ideas about biasing in the galaxy distribution.
\item A complete survey of high-mass galaxies, plus Tully-Fisher
estimates of distances and hence peculiar velocities, will provide
strong constraints on the density parameter.
\end{enumerate}

{\Large \bf Acknowledgments} 

I would like to acknowledge the support of a Nuffield Foundation
Science Research Fellowship and the hospitality of the University of
Melbourne where the preparation of this paper was undertaken.  Thanks
also to David Weinberg for supplying me with a copy of
Figure~\ref{fig:nhi}.

\bigskip
\reference Bond J. R., Cole S., Efstathiou G., Kaiser N., 1992, 
 ApJ, 379, 440
\reference Briggs F. H., 1990, AJ, 100, 999
\reference Couchman H. M. P., Thomas P. A., Pearce, F. R., 1995, 452, 797
\reference Henning P. A., 1995, ApJ, 450, 578
\reference Hoffman G. L., Lewis B. M., Helou G., Salpeter E. E.,
Williams H. L., 1989, ApJS, 69, 65
\reference Katz N., Weinberg D. H., Hernquist L., Miralda-Escud\'e J.
M., 1995, preprint, astro-ph/9506106
\reference Kerr F. J., Henning P. A., 1987, 320, L99
\reference Lacey C. G., Cole S., 1994, MNRAS, 271, 676
\reference Press W. H., Schechter P. G., 1974, ApJ, 187, 425
\reference Rees M. J., Ostriker J. P., 1977, MNRAS, 179, 541
\reference Salpeter E. E., Dickey, J. M., 1985, ApJ, 292, 426
\reference Schneider S. E. \etal, 1989, AJ, 97, 666
\reference Silk, J., 1977, ApJ, 211, 638

\end{document}